\documentclass{jnmp03c}

\usepackage{amsmath}

\begin{document}

\setcounter{page}{65}
\renewcommand{\evenhead}{O M Kiselev}
\renewcommand{\oddhead}{Hard Loss of Stability in Painlev\'e-2
Equation}

\thispagestyle{empty}

\FistPageHead{1}{\pageref{kiselev-firstpage}--\pageref{kiselev-lastpage}}{Article}

\copyrightnote{2001}{O M Kiselev}

\Name{Hard Loss of Stability in Painlev\'e-2
Equation}\label{kiselev-firstpage}

\Author{O M KISELEV}

\Adress{Institute of Mathematics, Ufa Sci.
Centre of Russian Acad. of Sci.\\
 112, Chernyshevsky str., Ufa, 450000, Russia\\
E-mail: ok@imat.rb.ru}

\Date{Received March 3, 2000; Revised March 14, 2000;
Accepted October 28, 2000}

\begin{abstract}
\noindent A special asymptotic solution of the Painlev\'e-2
equation with small parameter is stu\-died. This solution has a
critical point $t_*$ corresponding to a bifurcation phenomenon.
When $t<t_*$ the constructed solution varies slowly and when
$t>t_*$ the solution oscillates very fast. We investigate the
transitional layer in detail and obtain a smooth asymptotic
solution, using a sequence of scaling and matching procedures.
\end{abstract}

\section{Introduction}

In this work a special asymptotic solution for the equation
Painlev\'e-2
\begin{equation}
\varepsilon^2u'' + 2u^3 +tu = 1 \label{p2}
\end{equation}
is constructed as $\varepsilon\to0$.
\par
The behaviour of wanted solution differs in different intervals of
the parameter $t$. The qualitative behaviour of numerical solution
is indicated in the figure (at $\varepsilon^2=0.1$). Calculations for
another small values of $\varepsilon$ give pictures like this. Let us
explain these numeric results  using asymptotic theory for small $\varepsilon$.

\begin{figure}[th]
\centerline{\epsfig{file=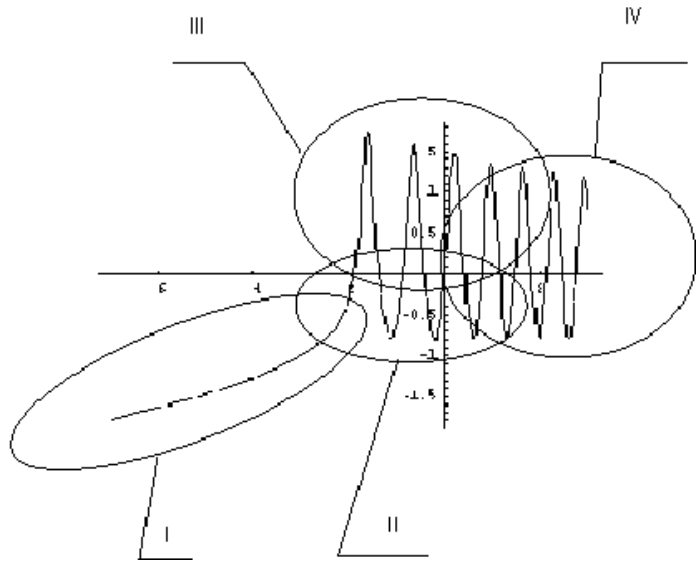,width=95mm}}
\caption{The hard loss of stability.}
\end{figure}

In the area I the special solution is approximated by an asymptotic
solution in which the leading term is a least root of a cubic
equation $2u^3+tu=1$. Corrections are algebraic functions of $t$.
Such algebraic asymptotics becomes invalid near the point $t_*$,
where two of the roots of the cubic equation coalesce. In the left
neighborhood of the point $t_*$ this asymptotics goes into an
asymptotics of a special solution of the Painlev\'e-1 equation with
respect to new scaling variable $\tau=(t-t_*)\varepsilon^{-4/5}$.
This is the area II in the figure. This special solution of the
Painlev\'e-1 equation has poles at $\tau=\tau_k$, $k=0,1,\dots$.  In
the neighborhoods of these poles one more scaling is done. New
variable is $\theta=(\tau-\tau_k)\varepsilon^{-1/5}$ (area~III). In
this area the asymptotics is defined by separatrix solution of a
nonlinear autonomous equation. This combined asymptotic structure
becomes invalid as $\tau\to\infty$, because the poles of the solution
Painlev\'e-1 equations close to each other. As $\tau\to\infty$ a
fast oscillating asymptotics is valid. It is the area IV in the
figure.

The qualitative behaviour of solutions of second-order ordinary
differential equations with respect to an additional parameter was
explained, for example, in the book \cite{Andr}. In~\cite{Andr} the
various types of bifurcations for equilibrium positions of
conservative second-order ordinary differential equations were
described also.

The equation  (\ref{p2}) is nonautonomous, however one can
separate the dependence on slow~$t$ and fast variables
($\tau,\theta$) in the asymptotic solution of this equation. We
may speak about the phase plane and phase trajectory of this
equation  with respect to  the fast variable $t/\varepsilon$. The
equilibrium positions on the  phase plane depend on $t$.  When
$t<t_*$ there are three equilibrium positions. At the critical
point $t_*$ the bifurcation ``saddle-center" occurs. It means that
one of stable and unstable equilibrium positions coalesce. When
$t>t_*$ only one equilibrium position exists. Such bifurcation
leads to instability~\cite{Arn}.

The bifurcations of slowly varying equilibrium positions of second
order equation with the algebraic nonlinearity and slowly varying
parameters were considered in \cite{Hab1} only in a~preliminary
fashion. When  this text was written, new work~\cite{Hab3} became
known. In~that work, a change of an energy and a phase jump has
been studied for a solution in a~very narrow layer near  a
saddle-center bifurcation point in general case. However results
of the work \cite{Hab3} are inapplicable to the Painlev\'e-2
equation, because this equation has degenerate behaviour with
respect to equations considered in \cite{Hab3}.

Amongst another works, in which the asymptotic solutions of the
nonlinear equations with varying coefficients were studied we
should note the work \cite{Neisht}. In that work a~chan\-ging of
an adiabatic invariant was studied in a problem when a solution
passed through separatrix in the nondegenerate case. One more work
where the passage through the separatrix in nondegenerate case was
studied is~\cite{Hab2}.

More complicated bifurcation is  the pitchfork in the equation
Painlev\'e-2. Asymptotics with respect to a small parameter of the
solutions for the equation Painlev\'e-2 with zero in the right hand
side of the equation (\ref{p2}) was investigated in works
\cite{Hab,Kar-Per}. In this case a solution in an interior
layer near a bifurcation value of the parameter $t_*$ is determined
by the equation Painlev\'e-2, but already without a small parameter;
and the problem, generally speaking, does not become simpler.

The asymptotics of the solutions for the Painlev\'e equations with a
leading term as an elliptic function with the modulated parameters
were studied, for example, in works \cite{Kap2}--\cite{V}. We must
mention some of the works about the scaling limits or double
asymptotics for the Painlev\'e-2 equation. The general approach to
the scaling limits of the Painlev\'e equations based on the
B\"acklund transformations was studied in \cite{Kit2}. The scaling
limit passage from the equation Painlev\'e-2 to Painlev\'e-1 was
studied on the level of classical solutions in the work
\cite{Kap-Kit}. In work \cite{Joshi} different approaches to the
double asymptotics were developed. The qualitative analysis for the
relation of the algebraic and fast oscillating asymptotic solutions
of the equation (\ref{p2}) was done in the work \cite{OK-BIS} also.
However, the asymptotic solutions constructed by this way are
non-uniform with respect to two variable $t$ and $\varepsilon$.

The major difference of the presented work in comparsion with the
others cited is constructing uniform asymptotic solution with
respect to two parameters $t$ and $\varepsilon$ as $\varepsilon\to 0$ into an
interval of $t$, where the  main term  of asymptotics (elliptic
function) is degenerated. This uniform asymptotic solution is valid
on a segment containing the saddle-center bifurcation point. In this
work we specify a different types of asymptotic approximations of
the being studied solution, their valid intervals and orders of
neglected terms which result  when these asymptotics are substituted
into a being solved equation.

The contents of the various sections are as follows. Section~2
states the main problem. The results are formulated in Section~3. An
asymptotic expansion to be valid before the bifurcation point $t_*$
is written in Section~4. Section~5 is devoted to inner asymptotic
expansions which are matched with each other and with asymptotics
from Section~4. A~fast oscillating asymptotic expansion of
Kuzmak-type, a degeneration of the oscillations and the matching of
the fast oscillations with inner expansions from Section~5 are
described in Section~6. Open problems are discussed in Section~7.

\section {Naive statement of the problem}

Let's consider a cubic equation:
\begin{equation}
2u^3 + tu  = 1,
\label{kub}
\end{equation}
which is obtained as a rejection of the
term with the small parameter in the equation (\ref {p2}).
There exist  the point $t_*$ and the value $u_*$ such, that
if $t=t_*$, then $u_*$ is double root of the equation
(\ref{kub}). The values $u_*$ and $t_*$ are easy to obtain
by solving of the equations:
\[
2u_*^3 + t_*u_* = 1, \qquad 6u_*^2 + t_*=0.
\]
There are  $t_*=-3\, 2^{-1/3}$, $u_*=-4^{-1/3}$.

The discriminant of the equation (\ref{kub}) has the form:
\[
D=\left(\frac t6\right)^3+\left(\frac 14\right)^2.
\]
The discriminant $D<0$ when $t<t_*$ and hence the cubic equation
(\ref{kub}) has three real roots $u_1(t)<u_2(t)<u_3(t)$. If
$t>t_*$ then $D>0$ and the cubic equation (\ref{kub}) has one
real root and two complex conjugate roots.  At $t=t_*$ the roots
$u_1(t)$ and $u_2(t)$ coalesce $u_1(t_*)=u_2(t_*)=u_*$.

When $t<t_*$ it is possible to construct a real formal
solution of the equation (\ref {p2}):
\begin{equation}
u(t,\varepsilon)=\sum_{k=0}^{\infty} \varepsilon^{2k}\overset{2k}{u}(t)
\label{form}
\end{equation}
by taking any of roots $u_j(t)$ as a leading term
$\overset{0}{u}(t)$. These three formal solutions are slowly
varying equilibrium positions for the equation (\ref{p2}).

Consider an equation
\begin{equation}
\varepsilon^2v'' + \left(6u_j^2(t)+t\right)v = 0, \qquad j = 1,2,3,
\label{l1}
\end{equation}
which describe small perturbations of the leading term $u_j(t)$ of
the asymptotic expansions~(\ref{form}). If $j=2$, then
$6u_2^2(t)+t<0$, so that the equation (\ref{l1}) has one exponentially
growing solution and one exponentially decreasing solution. Hence
corresponding asymptotic expansion is unstable with respect to small
perturbation of the leading term. Other\-wise if $j=1,\,3$ then the
coefficients in the equation~(\ref{l1}) are positive and the formal
solution~(\ref{form}) with $u_{j}(t)$ as the leading term of the
asymptotics is stable.

When  $t=t_*$ the roots $u_1(t)$ and $u_2(t)$ coalesce:
$u_1(t_*)=u_2(t_*)=u_*$ and when $t>t_*$, there exists  only one
equilibrium $u_3(t)$. Therefore $t_*$ is the bifurcation point for
the asymptotic solution (\ref{form}) in case the leading term
$\overset{0}{u}(t)\equiv u_1(t)$.

Our proposal is to construct a smooth asymptotic solution of the
equation (\ref{p2}) with the leading term $u_1(t)$ when $t<t_*$
on a segment $[t_*-a, t_*+a]$, $a=\mbox{const}>0$.

\section {The main results}

Here we describe a smooth asymptotic solution constructed in this
work. Following by V~P~Maslov \cite{M-O} we will use  the words
``asymptotic solution with respect to
$\mbox{mod}\left(O\left(\varepsilon^\alpha\right)\right)$",
namely, a function is said to be an asymptotic solution of
$\mbox{mod}\left(O\left(\varepsilon^a\right)\right)$ of the
equation~(\ref{p2}) if after its substitution into this equation
the latter is satisfied up to the terms of the
order~$O\left(\varepsilon^a\right)$.

When $t_*-a\le t<t_*,$ $(a=\mbox{const}>0)$ and
$(t_*-t)\varepsilon^{-4/5}\gg 1$ the asymptotic solution with
respect to
$\mbox{mod}\left(O\left(\varepsilon^6\right)+O\left(\varepsilon^6\left(t-t_*\right)^{-13/2}
\right)\right)$ has the form
\begin{equation}
u(t,\varepsilon)=u_1(t)+\varepsilon^2\frac{-2tu_1(t)}{(6u_1^2(t)+t)^4}+
\varepsilon^4\overset{2}{u}(t).
\label{exp10}
\end{equation}
The last term of the formal asymptotic solution as $t\to
t_*-0$ can be written as:
\[
\overset{2}{u}(t)=O\left((t-t_*)^{-9/2}\right).
\]

When $|t-t_*|\ll 1$ the asymptotic solution is defined by
two different types of the formal asymptotic expansions.
The first one has the form
\begin{equation}
u(t,\varepsilon)=u_* +
\varepsilon^{2/5}\overset{0}{v}(\tau)+\varepsilon^{4/5}\overset{1}{v}(\tau).
\label {exp20}
\end{equation}
Here the variable $\tau$ is defined by the formula
$\tau=(t-t_*)\varepsilon^{-4/5}$, the function $\overset{0}{v}(\tau)$ is
defined as the solution of the equation Painlev\'e-1:
\[
\frac{d^2\overset{0}{v}(\tau)}{d\tau^2}+6u_*\overset{0}{v}{}^2+u_*\tau=0,
\]
with the pure algebraic
asymptotic behavior as $ \tau\to-\infty $:
\[
\overset{0}v(t)=-\sqrt{-\frac{\tau}{6}}+O\left(\tau^{-2}\right).
\]
The formula
(\ref{exp20}) is asymptotic solution with respect to
$\mbox{mod}\left(O\left(\varepsilon^{8/5}\tau^2\right)+
O\left(\varepsilon^{8/5}\right)\right)$
as  $1\ll-\tau\ll\varepsilon^{-4/5}$.

The function $\overset{0}{v}(\tau)$ has  poles of  second order at
some points $\tau_k$, $k=1,2,\ldots$ (see, e.g.~\cite{G-L}):
\[
\overset{0}{v}(\tau)=-\frac{1}{u_*(\tau-\tau_k)^2}+O\left(\tau_k(\tau-\tau_k)^2\right).
\]
Near the poles the last term of the asymptotic solution can be written
as
\[
 \overset{1}{v}(\tau)=O\left((\tau-\tau_k)^{-4}\right),
\qquad \mbox{as}\quad \tau\to\tau_k.
\]
The expansion (\ref{exp20}) is suitable at
$\varepsilon^{-1/5}|\tau-\tau_k|\gg1$. The formula (\ref{exp20}) is asymptotic
solution with respect to
$\mbox{mod}\left(O\left(\varepsilon^{8/5}\right)+O\left(\varepsilon^{8/5}\tau^2\right)
+O\left(\varepsilon^{8/5}\tau_k\left(\tau-\tau_k\right)^{-8}\right)\right)$.

As $\tau\to\infty$ the main term  of the asymptotics (\ref{exp20}) is
\begin{equation}
\overset{0}{v}(\tau,\varepsilon)=\sqrt{\tau}\wp(s,g_2,g_3)+O\left(\tau^{-\gamma}\right),
\qquad \gamma=\mbox{const}>0.
\label{exp201}
\end{equation}
Here
\[
s=\frac 45 \tau^{5/4}+ \overset{0}{\sigma}(\chi),\qquad
\mbox{where}\quad \chi=\varepsilon^{2/5} \frac 57\tau^{7/4},
\]
the phase shift $\overset{0}{\sigma}$ is defined in Section 5.4.1 by
a formula (\ref{shift1}). The parameter of the Weierstrass elliptic
function $g_2=-2u_*$. The second parameter $g_3$ is defined by a
solution of an equation (see, e.g.~\cite{Kit}):
\[
\mbox{Re}\, \int_\Gamma d\lambda  \;\omega=0,
\]
where $\Gamma$ is any circle on an algebraic curve $\omega^2=\lambda^3+
\lambda/2-g_3/4$.
The last term of the asymptotics (\ref{exp20}) has the form:
\[
\overset{1}{v}(\tau)=O(\tau)+
O\left(\frac{\tau}{(\tau-\tau_k)^4}\right)\qquad\mbox{as}\quad
\tau\to\infty\quad \mbox{and}\quad \tau\not=\tau_k.
\]
The formula (\ref{exp20}) is asymptotic solution with respect to
$\mbox{mod}\Bigl(O\left(\varepsilon^{8/5}\right)+
O\left(\varepsilon^{8/5}\tau^{3/2}\right)+$
$\left.O\left(\varepsilon^{8/5}\tau_k^{3/2}(\tau-\tau_k)^{-8}\right)\right)$.
The expansion (\ref{exp20}) is suitable as
$\tau\ll\varepsilon^{-4/5}$ and
$\varepsilon^{-1/5}|\tau-\tau_k|\tau_k^{-1/4}\gg1$.

The second one, which is valid in the neighborhoods $|\tau-\tau_k||\tau_
k|^{1/5}\ll1$ of poles $\tau_k$ of the function
$\overset{0}{v}(\tau)$, reads as
\begin{equation}
u(t,\varepsilon)=u_*+\overset{0}{w}(\theta_k)+\varepsilon^{4/5}\overset{1}{w}(\theta_k),
\label{exp30}
\end{equation}
where $\theta_k=(\tau-\tau_k)\varepsilon^{-1/5}+\varepsilon^{1/5}\overset{1}{\theta}_k$.
The phase shift $\overset{1}{\theta}_k$ is defined by  a formula
(\ref{shift2}). The function $\overset{0}{w}(\theta_k)$ is
defined by the formula:
\[
\overset{0}{w}(\theta_k)=-\frac{16u_*}{4 + 16u_*^2\theta_k^2}.
\]
The last term of the formal asymptotics (\ref{exp30}) at
$|\theta|\to\infty$ can be written as
\[
\overset{1}{w}(\theta_k)=O\left(\theta_k^2|\tau_k|\right).
\]
The formula (\ref{exp30}) is asymptotic solution with respect to
$\mbox{mod}\left(O\left(\varepsilon^{8/5}\right)+ O\left(\varepsilon^{8/5}\theta_k^4\tau_k^2\right)
\right)$.

When $(t-t_*)\varepsilon^{-2/3}\gg1$ and $t<t_*+a$ the asymptotic
solution with  respect to $\mbox{mod}\left(O\left(\varepsilon^2\right)+
O\left(\varepsilon^2(t-t_*)^{-3}\right)\right)$
has the fast oscillating behavior:
\begin{equation}
u(t,\varepsilon)=\overset{0}{U}(t_1,t)+\varepsilon\overset{1}{U}(t_1,t).
\label{exp40}
\end{equation}
Here the last term of the asymptotics (\ref{exp40}) at
$t\to t_*+0$ can be written as
\[
\overset{1}{U}(t_1,t)=O\left((t-t_*)^{-3/2}\right).
\]
The leading term of the asymptotic solution satisfies the
Cauchy problem:
\[
(S')^2\left(\p_{t_1}\overset{0}{U}\right)^2=-\overset{0}{U}{}^4-
t\overset{0}{U}{}^2+2\overset{0}{U}+E(t),\qquad
\overset{0}{U}|_{t_1=0}=u_*.
\]
Here $t_1 = S(t)/\varepsilon+\phi(t)$. The function $E(t)$ is defined by
the equation
\[
I_0\equiv2\int_{\beta(t)}^{\alpha(t)}\sqrt{-x^4-tx^2+2x+E(t)}\, dx=2\pi,
\]
where $\alpha(t)$ and $\beta(t)$ ($\alpha(t)>\beta(t)$) are two real roots of the
equation $-x^4-tx^2 + 2x + E(t)=0$, other roots of this equation are
complex.

The phase function $S(t)$ is the solution of the Cauchy problem:
\[
T=S'\sqrt{2}\int_{\beta(t)}^{\alpha(t)}\frac{dx}{\sqrt{-x^4-tx^2+2x+E(t)}},
\qquad S|_{t=t_*}=0.
\]
Where $T$ is the constant defined by the formula
\[
T=\frac{\sqrt{2}C_*(k)}{2|u_*|^{1/2}}\left(\frac{3}{6-2k^2}\right)^{1/4},
\]
where $k\approx 0.463$ is the unique solution of the equation
\[
\int_0^\infty
dy\frac{-ky+k^2+1}{\left[(y-k)^2+1\right]^{5/2}}y^{5/2}=0,
\]
and
\[
C_*(k)=\int_0^\infty\frac{dy}{\sqrt{y\left[(y-k)^2+1\right]}}.
\]
The phase shift $\phi(t)$ is defined by an equation (see \cite{B-H}):
\[
\frac{\p_E I_0}{\p_E S'}\,\phi'=a=\mbox{const}.
\]

\noindent
{\bf Remark 1.} In this work the constant $a$ remains out side of
our analysis. Its value  may be defined by using the
monodromy-preserve method for the Painlev\'e-2 equation~\cite{F-N}.

\medskip

\noindent {\bf Remark 2.}  The domains of validity of the asymptotic
solution (\ref{exp10}) and the asymptotic solution (\ref{exp20})
intersect, so that these expansions match. The solution of the
Painlev\'e-1 equation which defines the asymptotics (\ref{exp20})
has infinite sequence of the poles $\tau_k$, $k=1,2,\dots.$ Near all
of these poles we match the asymptotic solutions~(\ref{exp20})
and~(\ref{exp30}). As the  number of  the pole $k\to\infty$ the
domain of validity of this complicated combine
asymptotics~(\ref{exp20}) and~(\ref{exp30}) intersects with the
domain of validity for the fast oscillating asymptotic
solution~(\ref{exp40}). Its allows to match the sandwiched
asymptotics with the fast oscillating asymptotic solution.

\section{The outer algebraic asymptotics}

The algebraic asymptotic solution (\ref{exp10}) of the equation
(\ref{p2}) is constructed here. This asymptotic solution is
suitable when $t<t_*$ and asymptotic behavior of  this solution
is investigated as $ t\to t_*-0$.

We construct the asymptotic solution of the equation
(\ref{p2}) as:
\begin{equation}
u(t,\varepsilon)=
\overset{0}{u}(t)+\varepsilon^2\overset{1}{u}(t)+\varepsilon^4\overset{2}{u}(t)
+ \cdots.
\label{exp1}
\end{equation}

Let's formulate the result of this section. The asymptotic solution
(\ref{exp10}) with respect to $\mbox{mod}\left(O\left(\varepsilon^6(t-t_*)^{-13/2}\right)\right)$,
where
$\overset{0}{u}(t)\equiv u_1(t)$ is least of the solutions of the
equation~(\ref{kub}), is suitable when $ (t_*-t)\varepsilon^{-4/5}\gg1$ and
$t>t_*-a,$ where $a=\mbox{const}>0$.

\subsection{Constructing the algebraic asymptotic solution}

Let's obtain the coefficients of the asymptotics (\ref{exp1}).
Substituting the ansatz (\ref{exp1}) into the equation (\ref{p2})
and  equating coefficients at identical powers of $\varepsilon$ we find
the sequence of the formulas for $\overset{k}{u}(t)$,
$k=0,1,2,\dots$.
\[
2\overset{0}{u}{}^3(t)+t\overset{0}{u}(t)=1,\qquad
\left(6\overset{0}{u}{}^2(t)+t\right)\overset{1}{u}(t)=-\overset{0}{u}{}''(t),
\]
\[
\left(6\overset{0}{u}{}^2(t)+t\right)\overset{2}{u}(t)=
-6\overset{0}{u}(t)\overset{1}{u}{}^2(t)-\overset{1}{u}{}''(t).
\]

The cubic equation for $\overset{0}{u}(t)$ when $t<t_* $ has
three real roots $u_1(t)<u_2(t)<u_3(t)$. As the leading term of
asymptotic expansion (\ref{exp1}) we choose $u_1(t)$. The second
derivative of $\overset{0}{u}(t)$ has the form:
\[
\overset{0}{u}{}''=
\left(\frac{\overset{0}{u}}{6\overset{0}{u}{}^2+t}\right)'=
\frac{2t\overset{0}{u}{}^2(t)}{\left(6\overset{0}{u}{}^2(t)+t\right)^3}.
\]
This allows to obtain the formula  for
$\overset{1}{u}(t)$.

It is easy to get the expressions for the following terms of the
asymptotic solution (\ref{exp1}). In an explicit form they are not
adduced here, however, it is important to note, that the power of the
denominator $\left(6\overset{0}{u}{}^2(t)+t\right)$ in the
coefficients of the asymptotics grows with each next step. The
$n$-th term of the asymptotic expansion as
$\left(6\overset{0}{u}{}^2(t)+t\right)\to 0$ has the form
\begin{equation}
\overset{n}{u}(t)=O\left(\left(6\overset{0}{u}{}^2(t)+t\right)^{-5n+1}\right).
\label{un}
\end{equation}

Let's write the asymptotic behavior  of the asymptotic expansion
(\ref{exp1}) as $t\to t_*$. For this purpose we shall calculate
the asymptotics of the expression $\left(6\overset{0}{u}{}^2(t)+t\right)$:
\[
\left(6\overset{0}{u}{}^2(t)+t\right)\Bigr|_{t\to
t_*}=-2u_*\sqrt{6}\sqrt{t_*-t}
+\frac{2}{3}(t_*-t)-\frac{5}{9\sqrt{6}u_*}(t-t_*)^{3/2}+O\left((t_*-t)^2\right).
\]
Using this formula and $\overset{0}{u}(t)$,
$\overset{1}{u}(t)$ we obtain:
\[
\ba{l}
\ds
u(t,\varepsilon)=u_*-\!\frac{1}{\sqrt6}\sqrt{t_*-t}+\frac{1}{18u_*}(t_*-t)
+
\varepsilon^2\!\left[-\frac{1}{3\,2^{10/3}}(t_*-t)^{-2}-O\left((t_*-t)^{-3/2}\right)\!\right]\!
\vspace{3mm}\\
\ds \phantom{u(t,\varepsilon)=}
+O\left(\varepsilon^4(t_*-t)^{-9/2}\right)+O\left((t_*-t)^{3/2}\right).
\ea
\]

\subsection{The domain of validity of the algebraic asymptotic solution}

The domain of validity for this expansion as $t\to t_*-0 $
is determined from the relation
$\varepsilon^2\overset{n+1}{u}(t)/\overset{n}{u}(t)\ll1$. It
follows from the formula (\ref{un}), that the expansion
(\ref{exp1}) is suitable when $(t_*-t)\varepsilon^{-4/5}\gg1$.

Evaluate the residual which is obtained when one substitutes the
asymptotic solution~(\ref{exp10}) into the equation (\ref{p2})
\[
F(t,\varepsilon)=-\varepsilon^6\left(\overset{2}{u}{}''+
2\overset{0}{u}\overset{1}{u}\overset{2}{u}+
6\overset{1}{u}{}^3\right) - \varepsilon^8
\left(6\overset{0}{u}\overset{2}{u}{}^2+
\overset{1}{u}{}^2\overset{2}{u}\right)-
\varepsilon^{12}\overset{2}{u}{}^3.
\]
Using the asymptotic behaviour of
the $\overset{k}{u}$,  $k=0,1,2$ as $t\to t_*-0$ one can obtain
\[
F(t,\varepsilon)=O\left(\varepsilon^6(t-t_*)^{-13/2}\right).
\]

\section{The inner asymptotics}

In this section the asymptotic expansions of solution of
(\ref{p2}) which are suitable in the small neighborhood of a
point $t_*$ are constructed. By following terminology of the
matching method \cite {I}, they are called ``the inner asymptotic
expansions".

\subsection {First inner expansion}

It follows from the consideration of the validity of the outer
expansion, made in the previous section, that it is natural to make
the following scaling of variables:
\[
(u-u_*)=\varepsilon^{2/5}v, \qquad (t-t_*)=\varepsilon^{4/5}\tau.
\]
As a result we write the equation (\ref{p2}) as
\begin{equation}
\frac{d^2v}{d\tau^2}+6u_*v^2 + u_*\tau=-\varepsilon^{2/5}\left(\tau v+2v^3\right).
\label{p1}
\end{equation}

In the limit as $\varepsilon\to0$ we obtain the equation Painlev\'e-1.
This asymptotic reduction is known as one of the scaling limits
for the Painlev\'e-2 equation \cite{Ince} (see also
\cite{Hab1,Kap}).

A solution of this equation has the asymptotic expansion  as
$\tau\to-\infty$:
\[
\ba{l}
\ds v(\tau,\varepsilon)=\left(-\sqrt{-\tau/6}+\frac{1}{48u_*\tau^2}+
\frac{49}{768 \sqrt{6}u_*^2(-\tau)^{9/2}}+\cdots\right)
\vspace{3mm}\\
\ds \phantom{v(\tau,\varepsilon)=} +
\varepsilon^{2/5}\left(-\frac{\tau}{18u_*}+\frac{1}{144\sqrt{6}(-\tau)^{3/2}}+
\cdots\right)+O\left(\varepsilon^{4/5}\tau^{3/2}\right).
\ea
\]

The asymptotic solution of the equation (\ref{p1}) we build as:
\begin{equation}
v(\tau,\varepsilon)=\overset{0}{v}(\tau)+\sum_{n=1}^{\infty}
\varepsilon^{2n/5}\overset{n}{v}(\tau), \label{exp2}
\end{equation}
where the function  $\overset{0}{v}(\tau)$ is the solution of the
Painlev\'e-1 equation.

Here it is shown, that the asymptotic solution (\ref{exp2}) is
suitable in the neighborhood of infinity (when $-\tau\ll\varepsilon^{-4/5}$)
and in the neighborhood of the poles for the function
$\overset{0}{v}(\tau)$: $ (\tau-\tau_k)\varepsilon^{-1/5}\gg1 $.

\subsubsection{Asymptotic behaviour as $\pbf{\tau\to-\infty}$}

The coefficients of the asymptotics are calculated from the
matching condition for the asymptotic expansion (\ref{exp1}) as
$t\to t_*$ and the expansion (\ref{exp2}) as $\tau\to-\infty$. In
particular, $\overset{0}{v}(\tau)$ has the algebraic asymptotics:
\begin{equation}
\overset{0}{v}(\tau)|_{\tau\to-\infty}=-\sqrt{-\tau/6}+
\frac{1}{48u_*\tau^2}+\frac{49}{768 \sqrt{6}u_*^2(-\tau)^{9/2}}+O\left(\tau^{-7}\right).
\label{asp11}
\end{equation}

In the book \cite{G-L} it is shown, that there exists the solution of
the Painlev\'e-1 equation with the asymptotics (\ref{asp11}). The
data of a monodromy for the solution of the Painlev\'e-1 equation
with the asymptotics (\ref{asp11}) are calculated in the work
\cite{Kap2}.

The first correction in the asymptotics (\ref{exp2}) satisfies the
equation
\begin{equation}
\frac{d^2\overset{1}{v}}{d\tau^2}+12u_*\overset{0}{v}\overset{1}{v}=
-\tau\overset{0}{v}-2\overset{0}{v}{}^3.
\label{lp1}
\end{equation}
The asymptotics of the solution for this equation as
$\tau\to-\infty$ has the form
\[
\overset{1}{v}(\tau)=-\frac{\tau}{18u_*}+\frac{1}{144\sqrt{6}(-\tau)^{3/2}}
+O\left(\tau^{-4}\right) .
\]

Asymptotics of the higher corrections is constructed by ordinary
way. The $n$-th correction as $\tau\to-\infty$ has an order:
\[
\overset{n}{v}(\tau)= O\left((-\tau)^{(n+1)/2}\right).
\]

\subsubsection{Validity of the asymptotic solution as $\pbf{\tau\to-\infty}$}

The requirement of validity for the asymptotics is
$\varepsilon^{2/5}\overset{1}{v}/\overset{0}{v}\ll1$. It reduces
to the condition $(-\tau)\ll\varepsilon^{-4/5}$.

The residual of the asymptotic solution (\ref{exp20}) has the form:
\[
F(\tau,\varepsilon)=-\varepsilon^{8/5}\left(6u_*\overset{1}{v}{}^2+\tau\overset{1}{v}+
6\overset{1}{v}\overset{0}{v}{}^2\right)-
\varepsilon^{10/5}\overset{0}{v}\overset{1}{v}{}^2-\varepsilon^{12/5}\overset{1}{v}{}^3.
\]
Using the asymptotic behaviour of $\overset{k}{v}$, $k=0,1$ as
$(-\tau)\ll\varepsilon^{-4/5}$ one can obtain
\[
F(\tau,\varepsilon)=O\left(\varepsilon^{8/5}\left(\tau^2\right)\right).
\]

\subsubsection{Asymptotic behaviour near the poles}

The function $\overset{0}{v}(\tau)$ has the poles when
$\tau\in(-\infty, \infty)$. Let's denote these poles by $\tau_k$. In
the neighborhood of the pole $\tau\to\tau_k\pm0$ the function
$\overset{0}{v}(\tau)$ is defined by the converging power series
(see e.g.~\cite{G-L})
\begin{equation}
\overset{0}{v}(\tau)=-\frac{1}{u_*(\tau-\tau_k)^2}+
\frac{\tau_k u_*}{10}(\tau-\tau_k)^2+
\frac{u_*}{6}(\tau-\tau_k)^3+c_k(\tau-\tau_k)^4+
O\left((\tau-\tau_k)^5\right).
\label{asp12}
\end{equation}
The constants $\tau_k$ and $c_k$ are the parameters of this solution. In
the review \cite{Kit} it is noted, that the problem on the connection
between the asymptotics of this solution at infinity and the constants
$\tau_k$ and $c_k$ is not investigated yet. The points of the poles
$\tau_k$ and appropriate constants $c_k$ can be obtained with the help
of the numerical calculation using the given asymptotics at infinity~(\ref{asp11}).

The asymptotics of $\overset{1}{v}$ as $\tau\to\tau_k\pm0$ may be
written as a sum of a certain solution of a nonhomogeneous
linearized Painlev\'e-1 equation $\overset{1}{v}_c(\tau)$
\[
\overset{1}{v}_c(\tau)=-\frac{1}{(\tau-\tau_k)^4}+\frac{\tau_k}{120u_*}
-\frac{1}{24u_*}(\tau-\tau_k)+\frac{9c_k}{10u_*^2}(\tau-\tau_k)^2
+O\left((\tau-\tau_k)^5\right),
\]
and two solutions of a homogeneous linearized equation $v_1(\tau)$,
$v_2(\tau)$:
\[
v_1(\tau)=\frac{1}{(\tau-\tau_k)^3}
+\frac{\tau_k u_*^2}{10}(\tau-\tau_k)
+\frac{u_*^2}{5}(\tau-\tau_k)^2 +2c_k
u_*(\tau-\tau_k)^3+O\left((\tau-\tau_k)^5\right),
\]
\[
v_2(\tau)=(\tau-\tau_k)^4+O\left((\tau-\tau_k)^8\right).
\]
Thus:
\begin{equation}
\overset{1}{v}=\overset{1}{v}_c(\tau)+\overset{1}{a}{}_k^\pm
v_1(\tau) + \overset{1}{b}{}_k^\pm v_2(\tau) . \label{aslp1}
\end{equation}
Here  $\overset{1}{a}{}_k^\pm$ and
$\overset{1}{b}{}_k^\pm$ are constants.

Higher corrections have the same form:
\[
\overset{n}{v}(\tau)=\overset{n}{v}_c(\tau)+\overset{n}{a}{}_k^\pm
v_1(\tau)+\overset{n}{b}{}_k^\pm v_2(\tau),
\]
where
\[
\overset{n}{v}_c(\tau)=O\left((\tau-\tau_k)^{-2(n+1)}\right), \qquad
\mbox{as}\quad \tau\to\tau_k.
\]

\subsubsection{Validity of the asymptotic solution as $\pbf{\tau\to\tau_k}$}

By using the asymptotics (\ref{asp12}) and (\ref{aslp1}) it is
easy to see, that the asymptotic expansion~(\ref{exp2}) is
suitable at
\[
\varepsilon^{-1/5}|\tau-\tau_k|\gg1.
\]

The residual of the asymptotic solution as $\tau\to\tau_k$ when
$\varepsilon^{-1/5}|\tau-\tau_k|\gg1$ is
\[
F(\tau,\varepsilon)=\varepsilon^{8/5}O\left(\frac{\tau_k}{(\tau-\tau_k)^4}\right)+
\varepsilon^{8/5}O\left((\tau-\tau_k)^{-8}\right).
\]

\subsection{Second inner expansion}

For the construction of the uniform asymptotics in the neighborhood
of the pole of the function $\overset{0}{v}$ it is necessary  to
make one more scaling of the independent variable and the function
(see \cite{Hab1}):
\[
(\tau-\tau_k)=\varepsilon^{1/5}\theta, \qquad\varepsilon^{-2/5}v=w.
\]
For function $w$ we obtain the equation:
\begin{equation}
\frac{d^2 w}{d\theta^2}+6u_*w^2+2w^3=-\varepsilon^{4/5}\tau_k(u_*+w)
-\varepsilon\theta(u_*+w).
\label{de3}
\end{equation}

The solution of this equation has following asymptotic expansion as
$\theta\to-\infty$:
\begin{equation}
\ba{l}
\ds w= -\frac{1}{u_*\theta^2} +\frac{1}{4u_*^3\theta^4}
+O\left(\theta^{-6}\right)+
\varepsilon^{1/5}\left(\frac{\overset{1}{a}{}_k^-}{\theta^3}+O\left(\theta^{-4}\right)\right)
\vspace{3mm}\\
\ds \phantom{w=} +\varepsilon^{2/5}\left(\left(\overset{1}{a}{}_k^-\right)^2
\frac{1}{u_*\theta^4}+O\left(\theta^{-6}\right)\right) +
\varepsilon^{3/5}\left(\overset{2}{a}{}_k^- \frac{1}{\theta^3}
+O\left(\theta^{-4}\right)\right)
\vspace{3mm}\\
\ds \phantom{w=} +
\varepsilon^{4/5}\left( -\frac{120\tau_k}{u_*}+\frac{\tau_k u_*}{10}\theta^2
+O\left(\theta^{-1}\right)\right)
\vspace{3mm}\\
\ds \phantom{w=}+
\varepsilon\left(\frac{u_*\theta^3}{6}+\frac{\theta}{24u_*}+\overset{1}{a}{}_k^-
\tau_k u_*^2\theta+O(1)\right)
\vspace{3mm}\\
\ds \phantom{w=} +
\varepsilon^{6/5}\left(\frac{9c_k}{10u_*^2}\theta^2+c_k\theta^4
+\frac{\overset{1}{a}{}_k^-
u_*^2}{5}\theta^2+O\left(\theta^1\right)\right)
+O\left(\varepsilon^{7/5}\theta^5\right)
\vspace{3mm}\\
\ds \phantom{w=}
+\varepsilon^{8/5}\left(
\overset{1}{b}{}_k^-\theta^4-\frac{\tau_k^2
u_*^3}{300}\theta^6+O\left(\theta^2\right)\right)
+O\left(\varepsilon^{9/5}\right).
\ea  \label{w-left-as}
\end{equation}
This long asymptotic formula shows, that  the constant
$\overset{1}{b}{}_k^-$ appears only in the correction of an
order $\varepsilon^{8/5}$. If we want to construct the first correction
of the asymptotics for the first inner expansion after the pole
$\tau_k$, we must construct the correction in order $\varepsilon^{8/5}$ for
the second inner expansion.

It is convenient to include a time shift depended on $\varepsilon$ into
the main term, and construct the asymptotic expansion depended on
a new time variable:
\[
\theta_k=\theta+\varepsilon^{1/5}\overset{1}{\theta}_k
+\varepsilon^{3/5}\overset{2}{\theta}_k,
\]
where $\overset{n}{\theta}_k=\mbox{const}$.

We search the asymptotic expansion for the solution of this
equation as a segment of an asymptotic series
\begin{equation}
w(\theta_k,\varepsilon)=\overset{0}{w}(\theta_k)
+\varepsilon^{4/5}\overset{1}{w}(\theta_k)+
\varepsilon\overset{2}{w}(\theta_k)
+\varepsilon^{6/5}\overset{3}{w}(\theta_k)+\varepsilon^{8/5}\overset{4}{w}(\theta_k).
\label{exp3}
\end{equation}

In this case the equation for the $w(\theta_k,\varepsilon)$ looks like:
\[
\ba{l}
\ds \frac{d^2 w}{d\theta_k^2}+6u_*w^2+2w^3=
-\varepsilon^{4/5}\tau_k(u_*+w)-\varepsilon\theta_k(u_*+w)
\vspace{2mm}\\
\ds \phantom{\frac{d^2 w}{d\theta_k^2}+6u_*w^2+2w^3=} +
\varepsilon^{6/5}\overset{1}\theta_k(u_*+w)
+\varepsilon^{8/5}\overset{2}{\theta}_k(u_*+w)+\cdots.
\ea
\]

It is shown here, that  the asymptotic solution (\ref{exp30}) is
the formal asymptotic solution of the equation (\ref{de3}) with
respect to $\mbox{mod}\left(O\left(\varepsilon^{8/5}\tau_k^2\theta^4\right)
+ O\left(\varepsilon^{9/5}\tau_k\theta^5\right)+
O\left(\varepsilon^2\theta^6\right)\right)$
when $|\theta\tau_k^{1/5}|\ll\varepsilon^{1/5}$.

The solution of the equation for the leading term of the
asymptotics (\ref{exp3}) is defined by the asymptotics as
$\tau\to\tau_k$ of the asymptotic expansion (\ref{exp2}), which is
outer with respect to (\ref{exp3}). This solution has the form
\begin{equation}
\overset{0}{w}(\theta_k)=-\frac{16u_*}{4+16u_*^2\theta_k^2}.
\label{sol}
\end{equation}

The constants $\overset{n}{\theta}_k$ are defined by asymptotics of
the function $w(\theta,\varepsilon)$. Using the formula~(\ref{w-left-as}) we
obtain:
\begin{equation}
\overset{n}{\theta}_k=\frac{u_*}{2}\overset{n}{a}{}_k^-,\qquad
n=1,2,\dots.
\label{shift2}
\end{equation}

The corrections in the expansion (\ref {exp3}) satisfy the linearized
equations
\[
\frac{d^2\overset{1}{w}}{d\theta_k^2}+
\left(12u_*\overset{0}{w}+6\overset{0}{w}{}^2\right)\overset{1}{w}=
-\tau_k\left(u_*+\overset{0}{w}\right),
\]
\[
\frac{d^2\overset{2}{w}}{d\theta_k^2}+
\left(12u_*\overset{0}{w}+6\overset{0}{w}{}^2\right)\overset{2}{w}=
\theta_k\left(u_*+\overset{0}{w}\right),
\]
\[
\frac{d^2\overset{3}{w}}{d\theta^2}+
\left(12u_*\overset{0}{w}+6\overset{0}{w}{}^2\right)\overset{3}{w}=
\overset{1}{\theta}_k\left(u_*+\overset{0}{w}\right),
\]
\[
\frac{d^2\overset{4}{w}}{d\theta^2}+
\left(12u_*\overset{0}{w}+6\overset{0}{w}{}^2\right)\overset{3}{w}=
-6\overset{1}{w}{}^2\left(\overset{0}{w}+u_*\right)+
\overset{2}{\theta}_k\left(u_*+\overset{0}{w}\right).
\]

The expression for $\overset{0}{w}$ can be used to obtain two
linearly independent solutions of the homogeneous equation for the
corrections:
\[
w_1=\frac{8\theta_k}{\left(1+4u_*^2\theta_k^2\right)^2},
\]
\[
w_2=\left[-\frac{1}{8}+2u_*^2\theta_k^2-u_*\theta_k^4+
\frac{2}{5}\theta_k^6+\frac{2u_*^2}{7}\theta_k^8\right]
\frac{1}{\left(1+4u_*^2\theta_k^2\right)^2}.
\]

By using these solutions of the homogeneous equation  it is easy
to get the solutions of the nonhomogeneous equations for the
corrections. The asymptotics of the corrections as
$\theta\to\infty$ has the form:
\[
\overset{1}{w}=\frac{\tau_ku_*}{10}\theta_k^2+
\frac{\tau_k}{120u_*}+\frac{\tau_k}{120} \theta_k^{-2}
-\frac{1}{160u_*^2} \theta_k^{-4}+O\left(\theta_k^{-6}\right),
\]
\[
\overset{2}{w}=\frac{u_*}{6}\theta_k^3
+\frac{1}{24u_*}\theta_k+O\left(\theta_k^{-5}\right),
\]
\[
\overset{3}{w}=\frac{1}{56
u_*^2}c_k\theta_k^4+O\left(\theta_k^2\right),
\]
\[
\overset{4}{w}=-\frac{u_*^3 \tau_0^2}{300}\theta_k^6+\left(
\frac{1}{56 u_*^2}\overset{k}{b}{}_k^-
- \frac{11 u_* \tau_k^2}{2100}\right)\theta_k^4+O\left(\theta_k^2\right).
\]

It is important to note that the leading term of the asymptotics of
$\overset{3}{w}$ as $\theta_k\to\pm\infty$ is the same. This term
defines the constants $c_k$ and hence the solution of the
Painlev\'e-1 equation before and after the pole. We have the same
value of the constant $c_k$ as $\theta_k\to-\infty$ and $\theta_k\to\infty$
and then we have the same asymptotic solution of Painlev\'e-1
equation before and after the pole $\tau_k$.

An opposite result takes place for the coefficient $\overset{4}{w}$
in order $\theta_k^4$ as $\theta_k\to\infty$. This coefficient is changed.
It is equal $\overset{1}{b}{}_k^-/\left(56 u_*^2\right)$ as
$\theta_k\to-\infty$ and $\overset{1}{b}{}_k^-/\left(56 u_*^2\right)-11
u_* \tau_k^2/2100$ as $\theta_k\to\infty$.

\subsubsection{Validity of second internal asymptotic solution}

Using the asymptotics for the corrections and the leading term, we
obtain, that the expansion (\ref{exp3}) is suitable when
$|\theta\tau_k^{1/5}|\ll\varepsilon^{-1/5}$. On the other hand,
the expansion (\ref{exp2}) is suitable at $|\theta|\gg1$. Hence,
the domains of the applicability for the expansions (\ref{exp2})
and~(\ref{exp3}) are intersected at realization of the condition
$|\tau_k|\ll\varepsilon^{-1}$. If we take into account also the
requirement of fitness of the asymptotic expansion (\ref{exp3}),
then get the restriction $|\tau_k|\ll\varepsilon^{-4/5}$. From
this inequality it follows, that in this section the formal
asymptotic expansions to be suitable when $|t-t_*|\ll1$ are
constructed.

We calculate the residual of second internal asymptotic expansion
using the asymptotic behaviour of $\overset{k}{w}$,
$k=0,1,2,3,4$:
\begin{equation}
 F(\theta,\tau_k,\varepsilon)=\varepsilon^{9/5}O\left(\tau_k\theta^5\right),
\label{res-int2}
\end{equation}
when $|\theta||\tau_k|^{1/5}\ll\varepsilon^{-1/5}$.

\subsection{Dynamics in the internal layer}

Using the asymptotic expansion of the second inner expansion as
$\theta\to\infty$ and the first inner expansion as $\tau \to\tau_k+0$ we
find that the first inner expansion after the pole has the form
(\ref{exp2}) where
\[
\overset{n}{v}(\tau)=\overset{n}{v}_c(\tau)+
\overset{n}{a}{}_k^+ v_1(\tau)+
\overset{n}{b}{}_k^+ v_2(\tau).
\]
Here
\[
\overset{n}{a}{}_k^+=\overset{n}{a}{}_k^-,\qquad
\overset{n}{b}{}_k^+=\overset{n}{b}{}_k^-
+\overset{n}{\Delta}_k, \qquad n=1,2.
\]
The shift $\overset{n}{\Delta}_k$ may be calculated from the
asymptotics of the second inner expansion as $\theta\to\infty$, for
$n=1$ we have obtained:
\[
\overset{1}{\Delta}_k=- \frac{22 u_*^3\tau_k^2}{75}.
\]

Thus  a behaviour of the asymptotic solution in the internal layer
is combined by the first and the second inner asymptotic
expansions.

\subsection{The asymptotics of the inner expansions as $\pbf{\tau\to\infty}$}

In the above sections we demonstrate the asymptotic behaviour of the
asymptotic  solution at $\tau\to-\infty$ and near the poles of the
solution for the Painlev\'e-1 equation. Below we study the asymptotic
behaviour of the solution as $\tau\to\infty$.

The regular asymptotic expansion on $\varepsilon$ is constructed in the
previous section concerning the first inner asymptotic expansion.
This asymptotics is not valid as large $\tau$. To use the first
inner asymptotic expansion as $\tau\to\infty$ we must modulate the
parameters of the solution of the Painlev\'e-1 equation and cancel
secular terms in the first and the second corrections in the
asymptotic expansion (\ref{exp2}) using singular perturbation
theory. Instead of modulating the parameters of the solution of
the Painlev\'e-1 equation it is more convenient to study the
modulation equation for parameters of the main term of the
asymptotics for the solution of the Painlev\'e-1 equation as
$\tau\to\infty$. This study will be developed in this subsection.

\subsubsection{Asymptotic behaviour of first inner expansion}

The elliptic asymptotics of the solution for the Painlev\'e-1
equation as $\tau\to\infty$ was obtained by P~Boutroux~\cite{B}.
Here we are interested in a connection formula for the solution of
the Painlev\'e-1 equation.  Namely we have the asymptotic behaviour
of the solution as $\tau\to-\infty$ and we need the asymptotic
behaviour of the same solution as $\tau\to\infty$. The Painlev\'e-1
equation is integrable by the monodromy-preserving method
\cite{F-N}. If the monodromy data are known, then the solution of
the Painlev\'e-1 equation is uniquely defined. In the correspondence
with \cite{Kap2}, in our case the monodromy data are constants $s_2$
and $s_3$ and these constants are equal to zero. The asymptotics of
the function $\overset{0}{v}(\tau)$ outside of the poles has the
form (see e.g.~\cite{Kit})
\begin{equation}
\overset{0}{v}=\sqrt{\tau}\;\overset{0}{\rho}(\sigma)+O\left(\tau^{-\gamma
}\right), \label{asP1}
\end{equation}
where $\gamma>0$ is some constant, the function
$\overset{0}{\rho}(\sigma)$ is determined by the Weierstrass elliptic
function
\begin{equation}
\overset{0}{\rho}(\sigma)=-{\wp}(\sigma,g_2,g_3)/u_*. \label{ro0}
\end{equation}
The phase function $\sigma=\frac{4}{5} \tau^{5/4}$. It is
important to note, that in the formula (\ref{asP1}) the shift of
the phase function $\sigma$ is equal to zero. A parameter is
$g_2=-2u_*$ and a parameter $g_3$ is defined as a solution of the
equation:
\[
\mbox{Re} \int_\gamma \omega\,d\lambda=0,
\]
where $\gamma$ is any circle on an algebraic curve:
$\omega^2=\lambda^3+\lambda/2-g_3/4$.

We are interested in the asymptotic solution of the perturbed
Painlev\'e-1 equation (\ref{p1}). Therefore the asymptotics of
the works \cite{Kap,Kit} for unperturbed Painlev\'e-1
equation are the main term of our asymptotics on $\varepsilon$.
Substitute into (\ref{p1}):
\begin{equation}
v(\tau,\varepsilon)=\sqrt{\tau}\rho(\sigma,\varepsilon),\label{p1-as-elliptic}
\end{equation}
where $\sigma=4\tau^{5/4}/5$. As a result we obtain an equation:
\begin{equation}
\rho''+6u_*\rho^2+u_*=-\left(\frac{5\sigma}{4}\right)^{-1}\rho'+
\left(\frac{5\sigma}{4}\right)^{-2}\rho -
\varepsilon^{2/5}\left(\frac{5\sigma}{4}\right)^{2/5}\left(\rho+2\rho^3\right).
\label{W-p}
\end{equation}
Let us to construct the asymptotics of the perturbed equation as
a segment of the asymptotic series:
\begin{equation}
\rho(\sigma,\varepsilon)=\overset{0}{\rho}(s) +
\varepsilon^{2/5}\left(\frac{5\sigma}{4}\right)^{2/5}\overset{1}{\rho}(s)+
\varepsilon^{4/5}\left(\frac{5\sigma}{4}\right)^{4/5}\overset{2}{\rho}(s),
\label{W-as}
\end{equation}
where
\[
s=\sigma+\overset{0}{\sigma}(\chi),\qquad
\chi=\varepsilon^{2/5}\frac{5}{7}\left(\frac{5\sigma}{4}\right)^{7/5},
\]
$\overset{0}{\sigma}$ is modulated phase shift and $\chi$ is once more
slow variable.

Substitute the formula (\ref{W-as}) into the equation (\ref{W-p}).
Then let's equate the coefficients with identical powers of
$\varepsilon$. As a result we obtain the sequence of the equations:
\begin{equation}
\overset{0}{\rho}{}''+6u_*\overset{0}{\rho}{}^2 +
u_*=\left[-\left(\frac{5}{4}\sigma\right)^{-1}\overset{0}{\rho}{}'+
\frac{1}{4}\left(\frac{5}{4}\sigma\right)^{-2}\overset{0}{\rho}\right],
\label{W-e-t}
\end{equation}
\begin{equation}
\ba{l}
\ds
\overset{1}{\rho}{}''+12u_*\overset{0}{\rho}
\overset{1}{\rho}=-
\left(\overset{0}{\rho}+2\overset{0}{\rho}{}^3\right)-
\overset{0}{\sigma}{}'\overset{0}{\rho}{}''
\vspace{3mm}\\
\ds \phantom{\overset{1}{\rho}{}''+12u_*\overset{0}{\rho} \overset{1}{\rho}=}
+\left[-\frac{1}{2}\left(\frac{5}{4}\sigma\right)^{-1}
\overset{0}{\sigma}{}'\overset{0}{\rho}{}'
-2\left(\frac{5}{4}\sigma\right)^{-1}\overset{1}{\rho}{}' +
\frac{7}{8}\left(\frac{5}{4}\sigma\right)^{-2}\overset{1}{\rho} \right],
\ea \label{Lame1-t}
\end{equation}
\begin{equation}
\ba{l}
\ds \overset{2}{\rho}{}''+12u_*\overset{0}{\rho}\overset{2}{\rho}=
-6u_*\overset{1}{\rho}{}^2-\overset{1}{\rho}
\left(1+6\overset{0}{\rho}{}^2\right)-\left(\overset{0}{\sigma}{}'\right)^2\overset{0}{\rho}{}''-
\overset{0}{\sigma}{}''\overset{0}{\rho}'-\overset{0}{\sigma}{}'\overset{1}{\rho}{}''
\vspace{3mm}\\
\ds \phantom{\overset{2}{\rho}{}''+12u_*\overset{0}{\rho}\overset{2}{\rho}=}
 +  \left(\frac{5}{4}\sigma\right)^{-1}
L_1\left(\overset{0}{\rho},\overset{1}{\rho},\overset{2}{\rho}\right)+
\left(\frac{5}{4}\sigma\right)^{-2}
L_2\left(\overset{0}{\rho},\overset{1}{\rho},\overset{2}{\rho}\right).
\label{Lame2-t}
\ea
\end{equation}
Here $L_{1,2}$ are linear operators.

Solutions of these equations may be represented as asymptotic
expansions as $\sigma\to\infty$. We will assume that the corrections
of these expansions are small as $\sigma\to\infty$. Therefore  we will
neglect the correction terms of this asymptotics and consider
only the main terms of these asymptotics with respect to $\sigma$.
The equations for the main terms
$\overset{0}{\rho}_0$, $\overset{1}{\rho}_0$,
$\overset{2}{\rho}_0$ have the form:
\begin{equation}
\overset{0}{\rho}{}_0''+6u_*\overset{0}{\rho}{}_0^2 + u_*=0,
\label{W-e}
\end{equation}
\begin{equation}
\overset{1}{\rho}{}_0''+12u_*\overset{0}{\rho}_0
\overset{1}{\rho}_0=-
\left(\overset{0}{\rho}_0+2\overset{0}{\rho}{}_0^3\right)-
\overset{0}{\sigma}{}'\overset{0}{\rho}{}_0'',
\label{Lame1}
\end{equation}
\begin{equation}
\overset{2}{\rho}{}_0''+12u_*\overset{0}{\rho}_0\overset{2}{\rho}_0=
-6u_*\overset{1}{\rho}{}_0^2-\overset{1}{\rho}_0
\left(1+6\overset{0}{\rho}{}_0^2\right)-
\left(\overset{0}{\sigma}{}'\right)^2\overset{0}{\rho}{}_0''-
\overset{0}{\sigma}{}''\overset{0}{\rho}{}_0'
-\overset{0}{\sigma}{}'\overset{1}{\rho}{}_0''
\label{Lame2}
\end{equation}

The requirement of validity for the first inner asymptotics at
$\tau\to\infty$ is
$\varepsilon^{2n/5}\sigma^{2n/5}\overset{n}{\rho}/\overset{0}{\rho}\ll1$,
where $n=1,2.$  We find the modulated equation for the
$\overset{0}{\sigma}$ such that the first inner expansion is valid as
large $\tau$.

The solution of the equation for the main term is the function
$\overset{0}{\rho}(s)$ which is defined by (\ref{ro0}).  The
equations for the corrections is the Lame equations with external
force. To write the solutions of these equations we will use two
linear independent solutions of the Lame equation. Denote one of
these solutions by
\[
p_1(s)=\p_s\overset{0}{\rho}_0.
\]
We denote the second solution as $p_2(s)$. The solutions
$p_1(s)$ and $p_2(s)$ are such that a wronskian:
\[
W(p_1,p_2)=1.
\]
The second solution is aperiodic:
\[
p_2(s+\Omega)=C p_1(s)+p_2(s), \qquad \mbox{where} \quad C=\mbox{const}\not=0.
\]

The solution of the equation (\ref{Lame1}) can be written as:
\begin{equation}
\ba{l}
\ds \overset{1}{\rho}_0(s)=\overset{1}{A}_k p_1(s)+\overset{1}{B}_k
p_2(s)+ p_1(s)\int_{s_0}^{s}dz\left(-\overset{0}{\rho}_0(z)-
2\overset{0}{\rho}{}_0^3(z)\right)
p_2(z)
\vspace{3mm}\\
\ds \phantom{\overset{1}{\rho}_0(s)=} -
p_2(s)\int_{s_0}^{s}dz\left(-\overset{0}{\rho}_0(z)-
2\overset{0}{\rho}{}_0^3(z)\right)p_1(z) - \overset{0}{\sigma}{}'s
p_1(s).
\ea\label{rho1}
\end{equation}
Here $s_0=s_k+\Omega/2$, where $s_k$ is pole of the function
$\overset{0}{\rho}_0(s)$ and $\Omega$ is a real period of the~$\rho$.
The $\overset{1}{A}_k$ and $\overset{1}{B}_k$ are
constants.

The first correction $\overset{1}{\rho}$ is bounded as
$s\in{\mathbb R}$ if
\[
\overset{1}{B}_k=\frac{1}{C}\overset{0}{\sigma}{}'+
\frac{1}{C}\,R.P.\int_0^\Omega
dz\left(\overset{0}{\rho}_0(z)+2\overset{0}{\rho}{}_0^3(z)\right)p_1(z).
\]
In this formula the integral must be regularized. Namely:
\[
R.P.\int_0^\Omega
dz\left(\overset{0}{\rho}_0(z)+2\overset{0}{\rho}{}_0^3(z)\right)p_1(z)=
\mbox{res}_{r=0}\left[\frac{1}{r}
\int_r^{\Omega-r}dz\left(\overset{0}{\rho}_0(z)
+2\overset{0}{\rho}{}_0^3(z)\right)p_1(z) \right].
\]

Using a perturbation theory for the second order equation
developed in works \cite{Luke}--\cite{B-H}, we can show, that the
solution for the $\overset{2}{\rho}$ has an order $O(\sigma)$,
$\sigma\not=\sigma_k$, $k\in{\mathbb Z}$, if the
function~$\overset{0}{\sigma}$ is a solution of the differential
equation: $\overset{0}{\sigma}{}''=0.$ This equation and the
equation for the~$\overset{1}{B}_k$ allow to write the Cauchy
problem for the $\overset{0}{\sigma}$ in the form: \begin{equation} \ba{l} \ds
\overset{0}{\sigma}{}'=C\overset{1}{B}_k+{R.P.}\int_0^\Omega
dz\left(\overset{0}{\rho}_0(z)+2\overset{0}{\rho}{}_0^3(z)\right)p_2(z),
\vspace{3mm}\\
\ds  \chi\in\left(\chi_k,\chi_{k+1}\right),\qquad
\overset{0}{\sigma}|_{\chi=0}=0, \ea \label{shift1} \end{equation} where
\[
\chi_k=\varepsilon^{2/5}\frac{5}{7}\left(\frac{5\sigma_k}{4}\right)^{7/5}.
\]

The constants $\overset{1}{A}_k$ and $\overset{1}{B}_k$ are defined
by matching conditions of the asymptotics (\ref{p1-as-elliptic}),
(\ref{W-as}) and (\ref{exp2}) as $\tau_k\to\tau_k+0$:
\[
\overset{1}{A}_k=\frac{\overset{1}{a}{}_k^{+}}{\tau_k}-
{R.P.}\int_0^{\Omega/2}dz\left(\overset{0}{\rho}_0(z)+
2\overset{0}{\rho}{}_0^3(z)\right)p_2(z),
\]
\[
\overset{1}{B}_k=\frac{u_*\overset{1}{b}{}_k^+}{14\tau_k}-
{R.P.}\int_0^{\Omega/2}dz\left(\overset{0}{\rho}_0(z)
+2\overset{0}{\rho}{}_0^3(z)\right)p_1(z).
\]

As a result we obtain the formula (\ref{exp201}) for the main term of the
asymptotics (\ref{W-as}).

\subsubsection{Validity  of the first inner expansion as $\pbf{\tau\to\infty}$}

Let us denote by
\[
\overset{n}{v}(\tau)=\tau^{(n+1)/2}\overset{n}{\rho}(\sigma),\qquad
n=0,1,2.
\]
Then  the condition of the validity
$\varepsilon^{2/5}\overset{n}{v}/\overset{0}{v}\ll 1$, $n=1,2$
fulfills as $\varepsilon^{2/5}\tau/\sqrt{\tau}\ll1$ or as the same:
\[
\tau\ll\varepsilon^{-4/5}.
\]

Near the pole $\tau_n$ we obtain the asymptotics:
\[
 \overset{0}{v}=O\left(\frac{\sqrt{\tau}}{(\tau-\tau_k)^2}\right), \qquad
\mbox{and}\qquad
\overset{1}{v}=O\left(\frac{\tau}{(\tau-\tau_k)^4}\right).
\]
Hence the first inner asymptotic expansion is suitable near the
poles $\tau_k$ as
\[
\varepsilon^{-1/5}(\tau-\tau_k)\tau_k^{-1/4}\gg1.
\]

The residual of the first inner expansion is
\[
\ba{l}
F(\tau,\varepsilon)=-\varepsilon^{2}\left(12u_*\overset{1}{v}\overset{2}{v}-
6\overset{0}{v}\overset{1}{v}{}^2+6\overset{0}{v}{}^2\overset{2}{v}
+\tau\overset{2}{v}\right)
-\varepsilon^{12/5}\left(2\overset{1}{v}{}^3+
12\overset{0}{v}\overset{1}{v}\overset{2}{v}+
6u_*\overset{2}{v}{}^2\right)
\vspace{2mm}\\
\ds \phantom{F(\tau,\varepsilon)=} +
\varepsilon^{14/5}\left(6\overset{1}{v}{}^2\overset{2}{v}+
6\overset{0}{v}\overset{2}{v}{}^2\right)+
\varepsilon^{16/5}6\overset{1}{v}\overset{2}{v}{}^2+
\varepsilon^{18/5}2\overset{2}{v}{}^3.
\ea
\]

Using the results of the above section outside of the poles of
$\overset{0}{v}$ one can obtain:
\[
 F=O\left(\varepsilon^{2}\tau^{5/2}\right)+
O\left(\frac{\varepsilon^{2}\tau^{5/2}}{(\tau-\tau_k)^{10}}\right).
\]

\subsubsection{Validity of the second  inner expansion as $\pbf{\tau_k\to\infty}$}

In the second inner expansion the asymptotics as $\tau\to\infty$
corresponds to the asymptotics at $\tau_k\to\infty$. It is easy to
see, that in this case the first correction grows. This growth
limits the value $\tau_k$, at which the asymptotics (\ref{exp3}) is
correct: $|\tau_k|\ll\varepsilon^{-4/5}$. One can use the formula
(\ref{res-int2}) to  obtain the residual of the second inner
asymptotics as $\tau_k\to\infty$.

\section {Fast oscillating asymptotics}

\subsection {The Kuzmak's approximation}

In this section we apply  formulas obtained in
\cite{Kuz}--\cite{D-M}, \cite{B-H} to  the fast oscillating formal
asymptotic solution of the Painlev\'e-2 equation. These formulas are
usable when $t>t_*$.

The fast oscillating asymptotics is constructed as
\begin{equation}
u(t,\varepsilon)=\overset{0}{U}(t_1,t)+\varepsilon\overset{1}{U}(t_1,t) +
\varepsilon^2\overset{2}{U}(t_1,t)+\cdots.
\label{exp4}
\end{equation}
As the argument $t_1$ we use expression
\[
t_1=S(t)/\varepsilon+\phi(t),
\]
where $S(t)$ and $\phi(t)$ are unknown functions.

The equations for the leading term and  the
corrections of the asymptotics (\ref{exp4}) look like:
\begin{equation}
(S')^2\p_{t_1}^2\overset{0}{U}+2\overset{0}{U}{}^3+\overset{0}{U}t=1,
\label{eq-2U}
\end{equation}
\begin{equation}
(S')^2\p_{t_1}^2\overset{1}{U}+\left(6\overset{0}{U}{}^2+t\right)\overset{1}{U}=
-2S'\p^2_{tt_1}\overset{0}{U}-S''\p_{t_1}\overset{0}{U}-
2S'\phi'\p_{t_1}^2\overset{0}{U},
\label{eq-1U}
\end{equation}
\begin{equation}
\ba{l}
\ds
(S')^2\p_{t_1}^2\overset{2}{U}+\left(6\overset{0}{U}{}^2+t\right)\overset{2}{U}=
-6\overset{0}{U}\overset{1}{U}{}^2 - 2S'\p^2_{tt_1}\overset{1}{U}-
S''\p_{t_1}\overset{1}{U}-2S'\phi'\p_{t_1}^2\overset{1}{U}
\vspace{2mm}\\
\ds \phantom{(S')^2\p_{t_1}^2\overset{2}{U}+\left(6\overset{0}{U}{}^2+t\right)\overset{2}{U}=}
 - \p_t^2\overset{0}{U}-(\phi')^2\p_{t_1}\overset{0}{U}-
\phi''\p_{t_1}\overset{0}{U}-2\phi'\p_t\p_{t_1}\overset{0}{U}.
\ea \label{eq-U2}
\end{equation}

Integrating once with respect to $t_1$ the equation for
$\overset{0}{U}$ we obtain:
\begin{equation}
(S')^2\left(\p_{t_1}\overset{0}{U}\right)^2=-\overset{0}{U}{}^4-t\overset{0}{U}{}^2
+2\overset{0}{U}+E(t),
\label{eq-U}
\end{equation}
where $E(t)$ is the ``constant of integration".

We study the equation when the right-hand side has two real roots
$\beta(t)<\alpha(t)$. We define the initial data as
\[
\overset{0}{U}|_{t_1=0}=\beta(t).
\]
Solution of the equation (\ref{eq-U}) is an elliptic function. The
left hand side of the formula (\ref{eq-U}) is positive for real
functions  $S'$ and $\overset{0}{U}$. The term of highest order
in the right hand side is $-\overset{0}{U}{}^4$. Therefore the
real solution of the equation (\ref{eq-U}) has  no poles when
$t_1\in {\mathbb R}$ and  $|E(t)|<\infty$.

To construct an uniform asymptotics (\ref{exp4}) we choose the
unknown functions $S(t)$, $\phi(t)$  and  the ``constant of
integration" $E(t)$ by a special way. They must satisfy to
anti-resonant conditions for equations (\ref{eq-2U})--(\ref{eq-U2}).
 The first condition is the boundedness of the
right hand side of (\ref{eq-1U}) as $t_1\in{\mathbb R}$. It is
satisfied if a period of the oscillates of the function
$\overset{0}{U}(t_1,t)$ on~$t_1$ is a constant (see for
example~\cite{Kuz}):
\begin{equation}
T=\sqrt{2}S'\int_{\beta(t)}^{\alpha(t)}\frac{dx}{\sqrt{-x^4-tx^2+2x+E(t)}}.
\label{eq-E}
\end{equation}

The next condition is a boundedness of the first correction
$\overset{1}{U}(t_1,t)$ when $t_1\in{\mathbb R}$. It gives equation
defining a main term of an action $I_0$ (see,~\cite{Kuz}):
\[
I_0=S'\int_0^T\left[\p_{t_1}\overset{0}{U}(t_1,t)\right]^2dt_1=
\mbox{const}.
\]
Using the explicit expression
for the derivative with respect to $t_1$
we present this formula in a some other form:
\begin{equation}
I_0=2\int_{\beta(t)}^{\alpha(t)}\sqrt{-x^4-tx^2+2x+E(t)}\, dx=\mbox{const},
\label{eq-S}
\end{equation}
where $\alpha(t)$ and $\beta(t)$ are the solutions of the equation
$-x^4-tx^2+2x+E(t)=0$. The equation (\ref{eq-S}) means that the
action is a constant. At the other hand-side the
equation~(\ref{eq-S}) defines the energy $E(t)$ as a function with
respect to slow time $t$.

At last the necessary condition of boundedness of the second
correction $\overset{2}{U}$ with respect to $t_1$ is the equation
(see, \cite{B-H}):
\begin{equation}
\frac{\p_E I_0}{\p_ES'}\phi'=a=\mbox{const}.
\label{eq-phi}
\end{equation}

The equations (\ref{eq-E})--(\ref{eq-phi}) define the parameters of
the main term of the asymptotics as functions with respect to $t$.
To find these functions one should define
corresponding constants.

Notice that to define the value of the action $I_0$ we construct
the asymptotic solution of the equation (\ref{p2}) when
$t>t_*$. Thus the polynomial of the fourth power on
$\overset{0}{U}$ in the right hand side of the equation
(\ref{eq-U}) can have no more than two various real roots
$\alpha(t)$ and $\beta(t)$. Hence this polynomial can be
submitted as:
\[
F(x,t)=(\alpha(t)-x)(x-\beta(t))\left((x-m(t))^2+n^2(t)\right).
\]

The degeneration of the elliptic integral at $t=t_*$ corresponds
to the case $m(t_*)=\beta(t_*)=u_*$ and $n(t_*)=0$. For this case it
is easy to calculate the constant in the right hand side of the
equation (\ref{eq-S}), which is equal to $2\pi$ (i.e. $I_0=2\pi$) and the
value of the parameter
$E(t_*)=E_*=\frac{4}{3}\left(\frac{1}{2}\right)^{2/3}$.

\subsection {Degeneration of the fast oscillating asymptotics}

In this subsections we calculate the asymptotic behaviour of the
phase functions $S(t)$  as $t\to t_*+0$.

The oscillating solution is degenerated as $t\to t_*+0$.
Let's construct the asymptotics of this solution in the
neighborhood of the degeneration point. For this purpose
we calculate the asymptotics of the phase function $S(t)$
and the function $E(t)$. Let's write the equation~(\ref{eq-S})~as:
\begin{equation}
\int^{\alpha}_{\beta}\sqrt{(\alpha-x)(x-\beta)\left[(x-m)^2+n^2\right]}\, dx=\pi,
\label{eq-E1}
\end{equation}
where $\alpha$, $\beta$, $m$, $n$ are real functions when $t\ge t_*$.
These functions satisfy the Vi\'eta equations:
\begin{equation}
\ba{l}
\alpha + \beta + 2m = 0,
\vspace{1mm}\\
\ds m^2 + n^2 + \alpha\beta + 2m(\alpha + \beta) = t,
\vspace{1mm}\\
\ds ( \alpha + \beta) \left(m^2 + n^2\right) + 2m\alpha\beta = 2,
\vspace{1mm}\\
\alpha\beta \left(m^2 + n^2\right)=-E.
\ea \label{vieta}
\end{equation}

The equation (\ref{eq-E}) and three equation from (\ref{vieta})
define the dependency $ \alpha$, $\beta$, $m$, $n $ on the parameter
$t$. The last equation in (\ref{vieta}) defines the function $E(t)$.
Let's make changes of variables: $E=E_*+g_1$, $t=t_*+\eta$,
$4m=m_*+m_1.$ After simple transformations of the equations
(\ref{vieta}) we obtain:
\begin{equation}
\ba{l}
\ds 2m_*\left[6m_1^2-2n^2+\eta\right]
+\left[2m^2_1-2n^2+\eta\right]2m_1=0,
\vspace{1mm}\\
\ds m_*^2\left(12m_1^2-4n^2+\eta\right)+2m_*m_1
\left(6m_1^2-2n^2+\eta\right)
\vspace{1mm}\\
\ds \qquad \qquad +
\left(3m^2_1-n^2+\eta\right)\left(m_1^2+n^2\right)=-g_1.
\label{v2}
\ea
\end{equation}
Construct the solution of this system as $t\to t_*+0$ as:
\[
m_1=\mu\sqrt{\eta}+O(\eta),\qquad
n=\nu_1\sqrt{\eta}+O(\eta),\qquad
g_1=\gamma_1\eta+O\left(\eta^{3/2}\right).
\]

Let's substitute these expressions in (\ref{v2}), equate the coefficients at the
identical powers of $\eta$. As a result we obtain:
\[
6\mu_1^2-2\nu_1^2=-1, \qquad
\gamma_1=m_*^2.
\]

To define the constants $\mu_1$ and $\nu_1$ it is
necessary to construct the asymptotics as $\eta\to +0$ of
the left hand side of the equation (\ref{eq-E1}). The
asymptotics of the outside the integral coefficient in the
equation (\ref{eq-E1}) has the form
\begin{equation}
(\alpha-\beta)^3=64|m|^3\left[1-\frac{3}{2}
\frac{\mu_1\sqrt{-\eta}}{m_*}+\frac{3}{2}
\frac{\nu_1^2-\mu_1^2-1}{4m_*^2}\eta+O\left(\eta^{3/2}\right)\right].
\label{a-b}
\end{equation}

The integral in the equation (\ref{eq-E1}) is presented as
\[
I(k,\delta)=\int_0^1 dz\,\sqrt{(1-z)z}\sqrt{(z-k\delta)^2+\delta^2},
\]
where
\begin{equation}
z=\frac{x-\beta}{\alpha-\beta}, \qquad
\frac{m-\beta}{\alpha-\beta}=k\delta, \qquad \delta^2=\frac{n^2}{(\alpha-\beta)^2},
\label{k1}
\end{equation}
The value of the constant $k$ will be defined from an asymptotics
below.

The asymptotics of an integral $I(k,\delta)$ as $\delta\to0$ has
the form
\begin{equation}
I(k,\delta)=\frac{\pi}{16}-k\delta\frac{\pi}{8}
+\delta^2\frac{\pi}{4}+c(k)\delta^{5/2}
+O\left(\delta^3\right),
\label{asI}
\end{equation}
where
\[
c(k)=-\frac{8}{5}\int_0^\infty dy\frac{-ky+k^2+1}{\left[(y-k)^2+1\right]^{5/2}}y^{5/2}.
\]

First three terms in this formula are calculated by standard way.
Let's show as we  can obtain the function $c(k)$. For this purpose
the following trick (\cite{Fed1}) is applicable. Let's calculate
third derivative with respect to $\delta$ of the function
$I(k,\delta)$:
\[
\frac{\p^3I}{\p\delta^3}=-3\int_0^1dz\,\sqrt{(1-z)z}
\frac{-kz+k^2\delta+\delta}{\left[(z-k\delta)^2+\delta\right]^{5/2}}.
\]
On the right hand side we replace $z$ by $\delta y$ and we present the
integral as
\begin{equation}
\frac{\p^3I}{\p\delta^3}=-3\delta^{-1/2}\int_0^{\infty}dy\,y^{5/2}
\frac{-ky+k^2+1}{\left[(y-k)^2+1\right]^{5/2}}+O(1).
\label{de1}
\end{equation}

Solving the ordinary differential equation (\ref{de1}) in the
neighborhood of $\delta=0$, we get:
\[
I(k,\delta)=c_0+\delta c_1+\delta^2c_2+\delta^{5/2}\frac{8}{15}c_3(k)+
O\left(\delta^3\right),
\]
where
\[
C_3(k)=-3\int_0^\infty\,du\frac{-ky+k^2+1}{\left[(y-k)^2+1\right]^{5/2}}y^{5/2}.
\]
After that it is easy to obtain the asymptotics (\ref{asI}).

To define the value of $k$ we substitute the asymptotics (\ref{a-b})
and (\ref{asI}) in~(\ref{eq-E1}) and equate to zero the coefficients
at identical powers of $\eta$. In the result we get at~$\eta^{5/4}$
the equation
\[
c(k)=0.
\]

This is the transcendental equation for the definition of the
parameter $k$. The numerical solution gives $k\sim0.463$. Using the
formula (\ref{k1}), we get:
\[
\mu_1=\frac{k}{3}|\nu|,\qquad \nu_1=\sqrt{\frac{3}{2\left(3-k^2\right)}}.
\]

To construct the asymptotics of $S(t)$ as $t\to t_*+0$ we use the
equation connecting the period of fast oscillations with its phase
(\cite{Kuz}):
\begin{equation}
T=\sqrt{2}S'\int_\beta^\alpha\frac{dx}{\sqrt{(\alpha-x)(x-\beta)\left[(x-m)^2+n^2\right]}}.
\label{eq-S2}
\end{equation}
Present the integral in the right hand side as
\[
J=\frac{1}{\alpha-\beta}\int_0^1\frac{dz}{\sqrt{(1-z)z\left[(z-k\delta)^2+\delta^2\right]}}.
\]
After the same replacements, as at the construction of the
asymptotics $\frac{\p^3I}{\p\delta^3}$, as $\delta\to0$ we get:
\[
J=\frac{\delta^{-1/2}}{\alpha-\beta}\int_0^{\infty}
\frac{dy}{\sqrt{y\left[(y+k)^2+1\right]}}
+O(1).
\]
We substitute this expression into the equation
(\ref{eq-S2}), use the asymptotics $\delta$ and $(\alpha-\beta)$ as
$\eta\to+0$ and in the result we get:
\[
S'=(t-t_*)^{1/4}S_*(k)+O\left((t-t_*)^{1/2}\right),
\]
where
\[
S_*(k)=\frac{T}{\sqrt2}\frac{2|m_*|^{1/2}}{C_*(k)}
\left(\frac{3}{6-2k^2}\right)^{1/4},
\qquad C_*(k)=\int_0^\infty\frac{dy}{\sqrt{y\left[(y-k)^2+1\right]}}.
\]

The period of the oscillations for the function
$\overset{0}{U}(t_1,t)$ with respect to the variable $t_1$ in the
Krylov--Bogolubov's method is an arbitrary constant. Let's choose it
such, that $S_*(k)=1$:
\begin{equation}
T=\frac{S_*(k)\sqrt{2}C_*(k)}{2|u_*|^{1/2}}
\left(\frac{3}{6-2k^2}\right)^{1/4}.
\label{T}
\end{equation}
In the result the phase of the oscillations as $t\to t_*$
has a form
\begin{equation}
S(t)=\frac{4}{5}(t-t_*)^{5/4}+O\left((t-t_*)^{3/2}\right)+S_0,
\label{phase}
\end{equation}
where $S_0$ is  some constant. Its value will be defined below at
the matching of the asymptotics (\ref{exp4}) and inner asymptotics
(\ref{exp2}), (\ref{exp3}) as $t\to t_*+0$.

\subsection{The domain of validity of the  fast oscillating
asymptotics}

In this subsection we establish the domain of validity of
the fast oscillating asymptotics and compute the residual
of this asymptotic solution.

The validity of the asymptotics is defined by the formula
$\varepsilon\overset{1}{U}\ll\overset{0}{U}$. Let us check this
requirement. For this we must obtain the order of the first
correction as $t\to t_*+0$. Evaluate the order of the right hand
side of the equation for the first correction:
\[
F_1(t_1,t,\varepsilon)=-2S'\p^2_{tt_1}\overset{0}{U}-S''\p_{t_1}\overset{0}{U}.
\]
From the equation for $\overset{0}{U}$ one can evaluate second term
in $F_1(t_1,t,\varepsilon)$ as $t\to t_*+0$:
\[
S''\p_{t_1}\overset{0}{U}=O\left((t-t_*)^{-1}\right).
\]

One must reduce formula for the derivative of $\overset{0}{U}$ with
respect to $t$ to evaluate of the first term in the formula for
$F_1(t_1,t,\varepsilon)$.

The function $\overset{0}{U}$ is the inverse function with  respect to the
elliptic integral
\[
t_1+t_0=S'\int_{\beta(t)}^{\overset{0}{U}}
\frac{dy}{\sqrt{-y^4-ty^2+2y+E(t)}}.
\]
Both limits of  the integration are functions with respect
to $t$, it is not convenient for us. Make the substitution
$y=(\alpha-\beta)z+\beta$. Then  we obtain
\[
t_1+t_0=\frac{S'}{(\alpha-\beta)}
\int_{0}^{\frac{\overset{0}{U}-\beta}{\alpha-\beta}}
\frac{dz}{\sqrt{z(1-z)}
\sqrt{(z-\gamma)^2+\delta^2}}.
\]
Now we differentiate this formula with respect to $t$, as a result we obtain the
formula for the $\p_t
\left[\frac{\overset{0}{U}-\beta}{\alpha-\beta}\right]$:
\[
\ba{l}
\ds \p_t
\left[\frac{\overset{0}{U}-\beta}{\alpha-\beta}\right] =
\frac{1}{(\alpha-\beta)S'}
\sqrt{\left(\alpha-\overset{0}{U}\right)\left(\overset{0}{U}-\beta\right)
\left(\left(\overset{0}{U}-m\right)^2+n^2\right)}
\vspace{3mm}\\
\ds \phantom{\p_t \left[\frac{\overset{0}{U}-\beta}{\alpha-\beta}\right] =}
\times
\left[
-\p_t\left(\frac{S'}{(\alpha-\beta)}\right)
\int_0^{\frac{\overset{0}{U}-\beta}{\alpha-\beta}}
\frac{dz}{\sqrt{z(1-z)\left((z-\gamma)^2+\delta^2\right)}}\right.
\vspace{3mm}\\
\ds \phantom{\p_t \left[\frac{\overset{0}{U}-\beta}{\alpha-\beta}\right] =}
\left. +
\frac{S'}{(\alpha-\beta)}
\int_0^{\frac{\overset{0}{U}-\beta}{\alpha-\beta}}
\frac{dz\,\delta\delta'-(z-\gamma)\gamma'}
{\sqrt{z(1-z)}\left((z-\gamma)^2+\delta^2\right)^{3/2}}\right],
\ea
\]
where $\gamma=\frac{m}{\alpha-b}$.

In the same way we can obtain the formula:
\[
\ba{l}
\ds \p_t
\left[\frac{\overset{0}{U}-\alpha}{\alpha-\beta}\right]
=\frac{1}{(\alpha-\beta)S'}
\sqrt{\left(\alpha-\overset{0}{U}\right)\left(\overset{0}{U}-\beta\right)
\left(\left(\overset{0}{U}-m\right)^2+n^2\right)}
\vspace{3mm}\\
\ds \phantom{\p_t \left[\frac{\overset{0}{U}-\beta}{\alpha-\beta}\right] =}
\times
\left[ -\p_t\left(\frac{S'}{(\alpha-\beta)}\right)
\int_{-1}^{\frac{\overset{0}{U}-\alpha}{\alpha-\beta}}
\frac{dz}{\sqrt{z(1-z)\left((z-\Gamma)^2+\delta^2\right)}}\right.
\vspace{3mm}\\
\ds \phantom{\p_t \left[\frac{\overset{0}{U}-\beta}{\alpha-\beta}\right] =}
\left.+ \frac{S'}{(\alpha-\beta)}
\int_0^{\frac{\overset{0}{U}-\alpha}{\alpha-\beta}}
\frac{dz\,\delta\delta'-(z-\Gamma)\Gamma'}
{\sqrt{z(1-z)}\left((z-\Gamma)^2+\delta^2\right)^{3/2}}\right],
\ea
\]
where $\Gamma=\frac{\alpha-m}{\alpha-\beta}$.

\newpage

These formulas will be useful when we will reduce the
formula for the second derivative of $\overset{0}{U}$
with respect to $t$.

The first derivative of $\overset{0}{U}$ with respect to $t$ has the form:
\[
\ba{l}
\ds \p_t \overset{0}{U}=(\alpha-\beta)
\left[\frac{\alpha'-\beta'}{(\alpha-\beta)^2}+
\p_t\left(\frac{-\beta}{\alpha-\beta}\right)\right]
\vspace{3mm}\\
\ds \phantom{\p_t \overset{0}{U}=}
+\frac{1}{(\alpha-\beta)S'}
\sqrt{\left(\alpha-\overset{0}{U}\right)\left(\overset{0}{U}-\beta\right)
\left(\left(\overset{0}{U}-m\right)^2+n^2\right)}
\vspace{3mm}\\
\ds \phantom{\p_t \overset{0}{U}=}
\times
\left[
-\p_t\left(\frac{S'}{(\alpha-\beta)}\right)
\int_0^{\frac{\overset{0}{U}-\beta}{\alpha-\beta}}
\frac{dz}{\sqrt{z(1-z)\left((z-\gamma)^2+\delta^2\right)}}\right.
\vspace{3mm}\\
\ds \phantom{\p_t \overset{0}{U}=}
\left.+\frac{S'}{(\alpha-\beta)}
\int_0^{\frac{\overset{0}{U}-\beta}{\alpha-\beta}}
\frac{dz\,(z-\gamma)\gamma'+\delta\delta'}
{\sqrt{z(1-z)}\left((z-\gamma)^2+\delta^2\right)^{3/2}}\right].
\ea
\]

Now we can evaluate the second derivative $\p_{tt_1}^2\overset{0}{U}$.
\[
\p_{tt_1}^2\overset{0}{U}=\p_t\left[
\frac{1}{S'}
\sqrt{\left(\alpha-\overset{0}{U}\right)\left(\overset{0}{U}-\beta\right)
\left(\left(\overset{0}{U}-m\right)^2+n^2\right)}\right].
\]
Using the formula for $\p_t\overset{0}{U}$ one can obtain as
$t\to t_*+0$:
\[
\p_{tt_1}^2\overset{0}{U}=O\left((t-t_*)^{-5/4}\right).
\]

This formula allows to evaluate the right hand side in the
equation (\ref{eq-1U}) as $t\to t_*+0$:
\[
F_1(t_1,t,\varepsilon)=O\left((t-t_*)^{-5/4}\right).
\]

The first correction is periodical function with respect to $t_1$.
One can derive the solution of the equation for the first
correction  using two linear independent solution of the equation
\[
(S')^2\p_{t_1}^2V+\left(6\overset{0}{U}{}^2+t\right)V=0.
\]
Here our goal is to write these solutions in the terms of $\overset{0}{U}$
because then we evaluate the order of derivatives of the first
correction of asymptotic solution (\ref{exp4}) using the formula
for $\p_t \overset{0}{U}$.

The first one is
\[
U_1(t_1,t,\varepsilon)\equiv\p_{t_1}\overset{0}{U}= \pm\frac{1}{S'}
\sqrt{-\overset{0}{U}{}^4-t\overset{0}{U}{}^2
+2\overset{0}{U}+E(t)}.
\]
Here the sign before the root is $+$, when
$\p_{t_1}\overset{0}{U}>0$ and vice versa.

\newpage

The second solution of the homogeneous linearized equation
for the first correction is
\[
U_2(t_1,t,\varepsilon)=\pm \frac{1}{S'}
\sqrt{-\overset{0}{U}{}^4-t\overset{0}{U}{}^2
+2\overset{0}{U}+E(t)}
\int_{t_0}^{t_1} \frac{d\sigma}
{-\overset{0}{U}{}^4-t\overset{0}{U}{}^2+2\overset{0}{U}+E(t)}.
\]

Integral in this formula must be regularized because
integrand has second order poles at points
$\overset{0}{U}=\alpha$ and $\overset{0}{U}=\beta$. One of the
possibile way of the regularization is done in~\cite{Fed}.
Here we will follow~\cite{Fed}.

Near the singular points $\sigma=t_1^{j}$ $j\in{\mathbb  Z}$
one must represent the integral as
\[
\ba{l}
\ds S'\int_{t_1^{j}-\lambda}^{t_1^{j}+\lambda}
\frac{d\sigma}{U_1^2(\sigma,t,\varepsilon)}=
-S'\int_{t_1^{j}-\lambda}^{t_1^{j}+\lambda}
\frac{1}{\p_{\sigma}U_1(\sigma,t,\varepsilon)}
d_{\sigma}\left(\frac{1}{U_1(\sigma,t,\varepsilon)}\right)
\vspace{3mm}\\
\ds \qquad\qquad
 = -\frac{S'}{U_1(\sigma,t,\varepsilon)\p_{\sigma}U_1(\sigma,t,\varepsilon)}
\bigg|_{t_1^{j}-\lambda}^{t_1^{j}+\lambda}-
S'\int_{t_1^{j}-\lambda}^{t_1^{j}+\lambda}
\frac{d\sigma\,\p_\sigma^2  U_1(\sigma,t,\varepsilon)}
{U_1(\sigma,t,\varepsilon)\left(\p_\sigma U_1(\sigma,t,\varepsilon)\right)^2}.
\ea
\]
The parameter $\lambda$ may be for example $\lambda=T/4$,
where $T$ is the period of
oscillations of the function $\overset{0}{U}$ with respect to $t_1$.

Change the second derivative of $U_1$ in the last integrand as
\[
\p_{t_1}^2 U_1=-\left(\frac{1}{S'}\right)^2\left(6\overset{0}{U}{}^2+t\right)U_1,
\]
and the first derivative of $U_1$ as
\[
\p_{t_1}U_1=S'\p_{t_1}^2\overset{0}{U}=
\frac{1}{S'}\left(1-t\overset{0}{U}-2\overset{0}{U}{}^3\right).
\]
As a result we obtain formula for regularization of the integral:
\[
\ba{l}
\ds S'\int_{t_1^{j}-\lambda}^{t_1^{j}+\lambda}
\frac{d\sigma}{U_1^2(\sigma,t,\varepsilon)}=
-\frac{(S')^3}{U_1(\sigma,t,\varepsilon)\left(1-t\overset{0}{U}(\sigma,t,\varepsilon)-
2\overset{0}{U}{}^3(\sigma,\tau,\varepsilon)\right)}
\bigg|_{\sigma=t_1^{j}-\lambda}^{\sigma=t_1^{j}+\lambda}
\vspace{3mm}\\
\ds \phantom{S'\int_{t_1^{j}-\lambda}^{t_1^{j}+\lambda}
\frac{d\sigma}{U_1^2(\sigma,t,\varepsilon)}=}
+ S' \int_{t_1^{j}-\lambda}^{t_1^{j}+\lambda}
\frac{d\sigma \left(6\overset{0}{U}{}^2(\sigma,t,\varepsilon)+t\right)}
{\left(1-t\overset{0}{U}(\sigma,t,\varepsilon)-2\overset{0}{U}{}^3(\sigma,t,\varepsilon)
\right)^2}.
\ea
\]

Using the functions $U_1$ and $U_2$ one can solve the
equation (\ref{eq-1U}) for the first correction of the
asymptotic solution (\ref{exp4}) and obtain the solution
the terms of $\overset{0}{U}$.

One can see the first correction $\overset{1}{U}$ has the order of
the right hand side of the equation for the first correction
multiplyed on $S'$. It means that
\[
\overset{1}{U}=O\left((t-t_*)^{-3/2}\right)
\]
as $t\to t_*+0$.

\newpage

This formula  allows to obtain the restriction for the
validity of the formal asymptotic solution (\ref{exp4}):
\[
(t-t_*)\varepsilon^{-2/3}\gg1.
\]

Evaluate the residual of the asymptotic solution (\ref{exp4}). For
this we must evaluate  the  function
\[
F(t_1,t,\varepsilon)=-\varepsilon^2\p_{t}^2\overset{0}{U}-\varepsilon^2\p_{t}
\left(\frac{1}{S'}\p_{t_1}\overset{1}{U}\right)-\varepsilon^2
\frac{1}{S'}\p_{t_1}\p_{t}\overset{1}{U}- \varepsilon^3\p_{t}^2\overset{1}{U}.
\]
For all $t\in(t_*,t_*+a]$ the order of $F(t_1,t,\varepsilon)=O\left(\varepsilon^2\right)$.
But the order of $F$ grows as $t\to t_*+0$, because the derivatives
with respect to $t$ have singularity at $t=t_*$. In the right hand
side of this formula there is only functions on the
$\overset{0}{U}$, then we can differentiate the right hand side for
deriving of the second derivation of $\overset{0}{U}$ with respect
to $t$. The formula for $\p_t^2\overset{0}{U}$ will be very large if we
will write it in these definitions but  now one can evaluate the
order of $\p_t^2\overset{0}{U}$ as $t\to t_*+0$ using the formulas
\[
S'=O\left((t-t_*)^{1/4}\right),\qquad O(\gamma)=O(\delta)=O\left((t-t_*)^{1/2}\right),\qquad
\mbox{as} \quad t\to t_*+0.
\]
As a result one obtain:
\[
\p_t^2\overset{0}{U}(t_1,t,\varepsilon)=O\left((t-t_*)^{-10/4}\right).
\]

Using the same formulas one can evaluate the order of the $F(t_1,t,\varepsilon)$ as
$t\to t_*+0$:
\[
F(t_1,t,\varepsilon)=O\left(\varepsilon^2(t-t_*)^{-11/4}\right)
+O\left(\varepsilon^3(t-t_*)^{-17/4}\right).
\]

\subsection{The matching of the fast oscillating asymptotic solution\\
and the inner asymptotics}

The matching of this asymptotics with the inner asymptotics
(\ref{exp2}) and (\ref{exp3}) is carried out. From the matching
condition for the phase function we obtain the initial condition
$S(t)|_{t=t_*}=0$.

Now we turn to the evaluation of the asymptotics for the function
$\overset{0}{U}$ as $t\to t_*+0$. The function $\overset{0}{U}$ may
be written in the implicit form:
\begin{equation}
t_1=-S'\int_{\overset{0}{U}}^{\alpha(t)}\frac{dx}{\sqrt{-x^4-tx^2+2x+E(t)}}.
\label{imU-0}
\end{equation}
We remind the parameter $t_1$ is equal to $\varepsilon^{-1}S(t)+\phi(t)$ and the
additional term $S_0$ is undefined in the function $S(t)$. This
term we define  in this subsection.

The  formula (\ref{imU-0}) allows us to obtain the main term of
asymptotics for the $\overset{0}{U}$ at $t\to t_*+0$. Denote
\[
\overset{0}{U}=u_*+W(t_1,t)
\]
and
\[
 x=u_*+y.
\]
Using
asymptotics of $E(t)$ and $\alpha(t)$ as $t\to t_*+0$ one can get
\[
\ba{l}
\ds t_1=-S'\int_{W}^{u_*+\alpha(t)}\frac{dy}{\sqrt{-y^4-4u_*y^3+O(t-t_*)}}
\vspace{3mm}\\
\ds \phantom{t_1}=
S'\left(\int_{W}^{4u_*}\frac{dy}{\sqrt{-y^4-4u_*y^3}}+O\left((t-t_*)^{1/2}\right)+
O\left(\frac{(t-t_*)}{W^{5/2}}\right)\right).
\ea
\]

This formula allows to write the asymptotic expansion of
$\overset{0}{U}$ as $t\to t_*+0$ in the form:
\begin{equation}
\overset{0}{U}(t_1,t)=W_0(t_1/S')+O\left((t-t_*)^{1/2}\right)+
O\left(\frac{(t-t_*)}{W^{5/2}}\right). \label{exp5}
\end{equation}
The main term of the asymptotics  is defined by formula:
\[
W_0(t_1/S')=-\frac{4u_*}{1+4u_*^2(t_1/S')^2}.
\]
This asymptotics
is applicable as $W_0^{5/2}\gg(t-t_*)$. The function
$\overset{0}{U}(t_1,t)$ is periodic with respect to $t_1$. It
means the asymptotic (\ref{exp5}) is applicable on some segments
of the interval $\varepsilon^{4/5}\ll(t-t_*)\ll1$. The argument  of the
function  $W_0$ in the neighborhood of some point $t_k$ as $t\to
t_*+0$  is:
\[
\ba{l}
\ds \left(\frac{t_1}{\varepsilon S'}\right)\bigg|_{t=t_k}\sim
\frac{S(t_k)+S'(t_k)(t-t_k)+
O\left(S''(t_k)(t-t_k)^2\right)}{S'(t_k)+O\left(S''(t_k)(t-t_k)\right)}
\vspace{3mm}\\
\ds \phantom{\left(\frac{t_1}{\varepsilon S'}\right)\bigg|_{t=t_k}}
\sim
\frac{t-t_k}{\varepsilon}+S_k+O\left(\frac{(t-t_k)^2}{\varepsilon t_k}\right),
\ea
\]
where $S_k=\varepsilon^{-1}S(t_k)/S'(t_k)$.

It is easy to see the argument of the function $W_0$ may be
represented as
\[
\frac{t_1}{\varepsilon S'}\sim\theta+S_k,
\]
where $S_k$ is some constants depending on $S_0$ and number $k$.
One  can see the main term of the asymptotics (\ref{exp5})
coincides up to shift $S_k$ with the main term of the second inner
asymptotic expansion which is the function
$\overset{0}{w}(\theta_k)$. It is easy to see for full
definition of the function $\overset{0}{U}(t_1,t)$ one must find
the phase shift $S_0$.

We defined the additional constant $S_0$ by matching the functions
$\overset{0}{U}(t_1,t)$ and the first inner asymptotic expansion.

The formula (\ref{exp5}) is suitable when
$|W_0(\theta)|^{5/2}\gg|t-t_*|$. When $W_0(\theta)$ is small, we
consider other asymptotic formula for the function
$\overset{0}{U}(t_1,t)$:
\begin{equation}
\overset{0}{U}(t_1,t)=u_*+\sqrt{t-t_*}{\cal
P}\left(\frac{S(t)}{\varepsilon} +\phi(t),t\right). \label{exp6}
\end{equation}

Substitute this formula to the second-order equation for the
function $\overset{0}{U}(t_1,t)$ (\ref{eq-2U}). Expand the
function ${\cal  P}\left(S(t)/\varepsilon +\phi(t),t\right)$ with respect to
the small parameter $(t-t_*)$. In a result the equation for the
main term  of  the asymptotic expansion  is
\[
p''+6u_*p^2+u_*=0.
\]

This equation coincides with the equation for the asymptotics of
the first correction of the first inner asymptotic expansion. The
boundary conditions for the function $p(S(t)/\varepsilon+\phi(t))$ is
obtained from the condition of the matching (\ref{exp6}) with the
asymptotics of the expansion (\ref{exp5}) as $|\theta|\to\infty$.
The additional constant $S_0$ in the formula (\ref{phase}) is
finally defined at the matching of the asymptotic expansions
(\ref{exp5}), (\ref {exp6}) with asymptotics of the inner
asymptotic expansions. This get: $S_0=0$.

\section{Open problems}

In this work the bifurcation of the slowly varying equilibrium of
the Painlev\'e-2  equation was studied by matching method on the
formal asymptotic approach. However it is necessary to note
two important problems which remind out of side of our
analysis.
\begin{itemize}
\topsep0mm
\partopsep0mm
\parsep0mm
\itemsep0mm
\item[1.] The phase shift of the oscillating asymptotic solution
is  undefined in our approach. Its definition demands much more
thin calculations for the corrections of the asymptotic formulas.
\item[2.] A problem of a justification of the remainder
term for the constructed asymptotic solution remains open as well.
\end{itemize}

\subsection*{Acknowledgements}

This work was supported by RFBR (00-01-00663,
00-15-96038) and INTAS (99-1068).

I am grateful to A~N~Belogrudov, S~G~Glebov, L~A~Kalyakin, V~Yu~Novokshenov
and B~I~Suleimanov for stimulating discussions and also
V~E~Adler for the help in the realization of the numerical
calculating. Also I would like to thank R~Haberman kindly sending
his (collaborate with D~C~Diminnie) work~\cite{Hab3}. I am
grateful to A~V~Kitaev for many valuable comments, which resulted
in substantial improvement of the original version of this work.

\label{kiselev-lastpage}
\end{document}